\documentclass[aps,prl,a4paper,twocolumn,amsmath,showpacs]{revtex4}

\usepackage{graphicx}

\begin{document}
\bibliographystyle{apsrev}

\title{Enhancement by polydispersity of the biaxial nematic
phase \\ in a mixture of hard rods and plates}

\author{Yuri Mart\'{\i}nez-Rat\'on}
\email{yuri@math.uc3m.es}
\author{Jos\'e A.~Cuesta}
\email{cuesta@math.uc3m.es}
\affiliation{Grupo Interdisciplinar de Sistemas Complicados
(GISC), Departamento de Matem\'aticas, Unversidad Carlos III 
de Madrid, Avda.~de la Universidad, 30, E--28911, Legan\'es,
Madrid, Spain}

\date{\today}

\begin{abstract}
The phase diagram of a polydisperse mixture of uniaxial
rod-like and plate-like hard parallelepipeds is determined for 
aspect ratios $\kappa=5$ and 15. All particles have equal
volume and polydispersity is introduced in a highly symmetric
way. The corresponding binary mixture is known to have a
biaxial phase for $\kappa=15$, but to be unstable against
demixing into two uniaxial nematics for $\kappa=5$. We find
that the phase diagram for $\kappa=15$ is qualitatively similar 
to that of the binary mixture, regardless the amount of
polydispersity, while for $\kappa=5$ a sufficient amount
of polydispersity stabilizes the biaxial phase. 
This provides some clues for the design of an experiment
in which this long searched biaxial phase could be observed.
%We give some hints as to how this long searched biaxial phase could 
%be experimentally observed thanks to this enhancement of its
%stability by polydispersity.
\end{abstract}

% 64.70.Md  Transitions in liquid crystals
% 64.75.+g  Solubility, segregation, and mixing; phase separation
% 61.20.Gy  Theory and models of liquid structure

\maketitle

Systems of anisotropic molecules with two symmetry axes may form
a biaxial nematic phase. In this phase molecules align preferentially
along two perpendicular axes. 
%The phase diagrams of these systems were
%predicted to have a very unusual critical point arising from the
%confluence of two second-order transition with a first-order one
%in a sharp cusp \cite{alben:1973a}. 
The biaxial phase is experimentally difficult to observe because
in systems with biaxial molecules it is preempted by smectic or 
solid phases.
This led Alben \cite{alben:1973b} to propose an alternative system 
which should behave similarly, but in which spatial ordering is
difficulted: a mixture of hard rods and plates. His
analysis of this system with a mean-field lattice model showed a
phase diagram with four phases: isotropic fluid (I), rod-like
nematic (N$^+$), plate-like nematic (N$^-$), and biaxial (B). The
latter separates the two nematics with second-order transition lines
when composition is varied from rod-rich to plate-rich. Upon 
increasing concentration, for a rod-rich (plate-rich) composition the
system first undergoes a first-order I--N$^+$ (I--N$^-$) transition
and then a second order N$^+$--B (N$^-$--B) transition. At the 
crossover there is simply a continuous I--B transition. Two continuous 
and two first-order transitions meet at this multicritical point.
At the conclusions of his work, Alben mentions that a N$^+$--N$^-$ 
phase separation might replace the B phase, but does not considers 
this possibility in his analysis. A similar phase behavior has later
been obtained by other authors using different models \cite{rabin:1982,
stroobants:1984,sokolova:1997,chrzanowska:1997}. They take into
account the effect of having free (rather than restricted to a
lattice) rotations and/or translations. From them one learns 
three main things: first of all, that Alben's phase diagram is
qualitatively correct; secondly, that there is a {\em symmetric mixture},
namely that with rods and plates having the same molecular volume
(hence parallel or perpendicular like particles have the same
excluded volume), for which the multicritical point appears more or 
less at equimolar composition; and thirdly, that the phase diagram 
is perfectly symmetric about equimolarity only if virial coefficients 
higher than the second are neglected. 

The first experimental observation of a B phase was obtained by
Yu and Saupe \cite{yu:1980} in a ternary lyotropic mixture of
potassium laureate, 1--decanol and water. This system forms
lamellar and cylindrical micelles. By varying composition and
concentration N$^{\pm}$ and B phases appear, in a configuration
that reproduces Alben's phase diagram around the multicritical point.
Although this was first considered an experimental realization of 
Alben's model, it was latter recognized that micelles can really 
change shape from rod-like to plate-like through biaxial forms
as we move in the phase diagram. 
A Landau theory for a system of shape-changing micelles 
\cite{toledano:1994} does in fact reproduce even the most peculiar 
features of Yu-Saupe's system (like reentrance in the isotropic 
phase upon increasing concentration). So the experimental 
observation of
the B phase in a mixture of rods and plates remains a open
problem.

At the same time van Roij and Mulder \cite{vanroij:1994} introduced
an new important element in play. They considered the rod-plate
mixture version of Zwanzig's model (uniaxial
parallelepipeds), as well as an expansion of the free energy
up to the second virial coefficient. Then they reanalysed the
phase diagram with respect to N$^+$--N$^-$ demixing for the
symmetric mixture, for which
the excluded volume between unlike particles is minimum (hence
phase separation is least favored). What they found is that
N$^+$--N$^-$ phase separation is more stable than the B
phase up to aspect ratios (long-to-short axis ratio) $\kappa
\approx 8.8$. Above that threshold the phase diagram is like
that predicted by Alben up to a certain concentration, where
phase separation again replaces the B phase. The window of
stability of the B phase is relatively narrow. The driving
mechanism behind this phase behavior is the larger excluded
volume between unlike particles in the B phase as compared
to that between like particles (the rod-plate excluded volume 
divided by the rod-rod one scales as $\kappa^{2/3}$ for large
aspect ratios $\kappa$ \cite{vanroij:1994}). When the gain in free 
volume compensates the loss in mixing entropy (and this strongly 
depends on concentration and composition) phase separation 
occurs. This phase behavior was later confirmed in simulations
of symmetric mixtures of prolate and oblate ellipsoids
\cite{camp:1997}, the only difference being the logical asymmetry
in the phase diagram of the latter (an important difference, 
though, because it gives rise to N$^-$--B coexistence).

There is a new recent experiment, this time performed on a
true colloidal rod-plate mixture \cite{vanderkooij:2000b}. 
Rod (plate) aspect ratio is $\kappa\approx 10$ (15) while
plate-to-rod volume ratio is 13:1. The mixture is therefore
far from being symmetric. This system shows isotropic (I) and
uniaxial nematic phases, as well as biphasic (I--N$^{\pm}$) and 
triphasic (I--N$^+$--N$^-$) coexistences, but no B phase at all
(there appear more phases at higher concentrations, but they
are spatially inhomogeneous \cite{wensink:2001}). This picture
is consistent with the large volume difference between
rods and plates, and can in fact be explained with a Parsons-Lee
density functional approximation \cite{wensink:2001}. This
experiment introduces, however, another element of unpredicted
effect on the system: polydispersity. Both rods and plates have
about 20--30\% polydispersity in their axis lengths. For this
particular system, as theory shows \cite{wensink:2001}, this does
not seem to have any observable qualitative effect other than
allowing for more than three phase coexistence (which is forbidden
by Gibbs's phase rule in a two-component system) at high
concentrations. But as we aim to show in 
this letter, polydispersity has a more drastic effect in the phase 
behavior of the symmetric rod-plate mixture.

To this purpose, we have extended van Roij and Mulder's model in
two ways: (i) including polydispersity in a highly symmetric
way, in order to minimize trivial excluded volume effects; and
(ii) using fundamental-measure theory \cite{cuesta:1997a,
cuesta:1997b} to model its free energy, which for homogeneous 
phases is equivalent to employing a $y$-expansion exact up to 
the third virial term (hence the phase diagram
is expected to be asymmetric). Thus our system consists
of a mixture of parallelepipeds of size $L\times D\times D$, all
of them with equal volume: $LD^2=1$. If we characterize anisotropy
by $L/D=\lambda$, then $L=\lambda^{2/3}$ and $D=\lambda^{-1/3}$.
Rods have $1<\lambda<\infty$ while plates have $0<\lambda<1$.
The composition of the mixture is then given by a (parent)
probability density
\begin{equation}
\begin{split}
p(\lambda)&=\lambda^{-1}\big[\zeta f(\lambda/\kappa)+(1-\zeta)
f(\lambda\kappa)\big], \\
f(z)&=K_0(\alpha)^{-1}\exp[-(\alpha/2)(z^2+z^{-2})], 
\end{split}
\label{eq:parent}
\end{equation}
where $K_{\nu}(\alpha)$ ($\alpha>0$) is the $\nu$th-order modified 
Bessel function, and $\kappa>1$. Function $f(z)$ is peaked around
$z\approx 1$, and it is wider the smaller is $\alpha$; so 
$f(\lambda/\kappa)$ is peaked around
$\lambda=\kappa$ and $f(\lambda\kappa)$ is peaked around $\lambda=
\kappa^{-1}$. The former represents a polydisperse distribution of rods
and the latter a distribution of plates, both of aspect ratio $\kappa$.
Parameter $0\le\zeta\le 1$ allows to tune the overall composition of the
mixture, since the molar fraction of the rods is given by
$x_r(\zeta)\equiv\int_1^{\infty}p(\lambda)\,d\lambda$ (and that of plates
by $x_p(\zeta)=1-x_r(\zeta)$, of course). 
The probability density $p(\lambda)$ is chosen so that
if $dg(\lambda,\zeta)\equiv p(\lambda)d\lambda$, then $|dg(\lambda,\zeta)|=
|dg(\lambda^{-1},1-\zeta)|$; in other words, the fraction of rods and plates
within the same interval of aspect ratios are in the proportion 
$x_r(\zeta):x_p(1-\zeta)$.
The moments of this distribution are given by
$\langle\lambda^m\rangle=K_{m/2}(\alpha)K_0(\alpha)^{-1}[\zeta\kappa^m+
(1-\zeta)\kappa^{-m}]$, explicitly showing the symmetry of the mixture.

Fundamental-measure approximation for a multicomponent mixture of hard
uniaxial parallelepipeds amounts to taking for the excess (over the 
ideal) free energy density (in $kT$ units) 
\cite{cuesta:1997a,cuesta:1997b},
\[
\begin{split}
&\Phi=-n_0\ln(1-n_3)+\frac{\sum_{\mu}n_1^{\mu}
n_2^{\mu}}{1-n_3}+\frac{\prod_{\mu}n_2^{\mu}}{(1-n_3)^2}, \\
&\begin{array}{ll}
\displaystyle n_0=\sum_{\mu,i}\rho_i^{\mu}\equiv\rho, &
\displaystyle n_1^{\mu}=\sum_i\left[\rho D_i+
\rho_i^{\mu}\left(L_i-D_i\right)\right], \\
\displaystyle n_3=\sum_{\mu,i}\rho_i^{\mu}L_iD_i^2, &
\displaystyle n_2^{\mu}=\sum_i\left[\rho L_i-
\rho_i^{\mu}\left(L_i-D_i\right)\right]D_i,
\end{array}
\end{split}
\]
$\rho_i^{\mu}$ denoting the number density of parallelepipeds of
species $i$ with their symmetry axes oriented along the direction 
$\mu(=x,y,z)$.
Specializing for our system, where species are labeled by the
continuous parameter $\lambda$,
\begin{equation}
\begin{split}
\Phi &=-\rho\ln(1-\rho)
+\frac{\sum_{\mu}\xi_-^{\mu}\xi_+^{\mu}}{1-\rho}
+\frac{\prod_{\mu}\xi_+^{\mu}}{(1-\rho)^2}, \\
\xi_{\pm}^{\mu} &\equiv \int_0^{\infty}\lambda^{\pm 1/3}
\left[\left(\lambda^{\mp 1}-1\right)\rho^{\mu}(\lambda)+
\rho(\lambda)\right]\,d\lambda,
\end{split}
\end{equation}
where $\rho(\lambda)=\sum_{\mu}\rho^{\mu}(\lambda)$.
The total free energy density is then given by
\begin{equation}
\frac{\beta F}{V}=\sum_{\mu}\int_0^{\infty}\rho^{\mu}(\lambda)[
\ln\rho^{\mu}(\lambda)-1]+\Phi
\label{eq:totalPhi}
\end{equation}

Details for determining phase equilibria in a system like
this have been reported elsewhere \cite{clarke:2000,sollich:2002}. In brief,
if at a given $\rho$ there is only one phase present, its 
equilibrium composition is determined by minimizing (\ref{eq:totalPhi})
with respect to $\rho^{\mu}(\lambda)$ under the constraint
$\rho(\lambda)=\rho p(\lambda)$, with $p(\lambda)$
given by (\ref{eq:parent}). This amounts to solving the system
of equations ($\mu=x,y,z$)
\begin{equation}
\rho^{\mu}(\lambda)=\rho p(\lambda)\frac{e^{-\Phi^{\mu}(\lambda)}}{
\sum_{\nu}e^{-\Phi^{\nu}(\lambda)}},\quad
\Phi^{\mu}(\lambda)\equiv\frac{\delta\Phi}{\delta\rho^{\mu}(\lambda)}.
\end{equation}
If $m>1$ phases are present in mutual equilibrium, then labeling
them with $a(=1,\dots,m)$ their density distributions must verify 
the `lever rule'
\begin{equation}
\rho p(\lambda)=\sum_av_a\rho_a(\lambda),
\end{equation}
$v_a$ denoting the fraction of volume ocupied by the $a$th phase
($\sum_av_a=1$). Chemical equilibrium then yields
\begin{equation}
\rho_a^{\mu}(\lambda)=\rho p(\lambda)\frac{e^{-\Phi_a^{\mu}(\lambda)}}{
\sum_bv_b\sum_{\nu}e^{-\Phi_b^{\nu}(\lambda)}},
\end{equation}
and this system of equations is completed by the equality of
osmotic pressures between every pair of phases, where osmotic
pressure is obtained from (\ref{eq:totalPhi}) as
\begin{equation}
\beta\Pi=\frac{\rho}{1-\rho}+\frac{\sum_{\mu}\xi_-^{\mu}\xi_+^{\mu}}{
(1-\rho)^2}+2\,\frac{\prod_{\mu}\xi_+^{\mu}}{(1-\rho)^3}
\end{equation}

First of all, as this theory has contributions of virial terms
higher than the second, we have checked van Roij and Mulder's
results \cite{vanroij:1994}. So we have first chosen a bidisperse
mixture (pure rods and pure plates). The resulting phase diagrams
are qualitatively the same, apart from some asymmetry induced by
those higher virials: for $\kappa=15$ there is a small B window bounded
above by N$^+$--N$^-$ demixing, whereas no B phase appears for
$\kappa=5$. In both cases, the multicritical point is found very 
near $x_r=1/2$. An important difference produced by the asymmetry is
the replacement of the very narrow region of triphasic coexistence 
found by van Roij and Mulder by a simple N$^+$--B coexistence.
It is also worth mentioning that the asymmetry we obtain mirrors
that of the simulations of prolate and oblate ellipsoids
\cite{camp:1997} (they observe N$^-$--B coexistence instead). This 
means that the asymmetry must be strongly influenced by the details 
of the model (mainly the shape of the molecules and the restriction
of orientations).

\begin{figure}
\hspace*{4mm}\mbox{\includegraphics*[width=3.0in, angle=0]{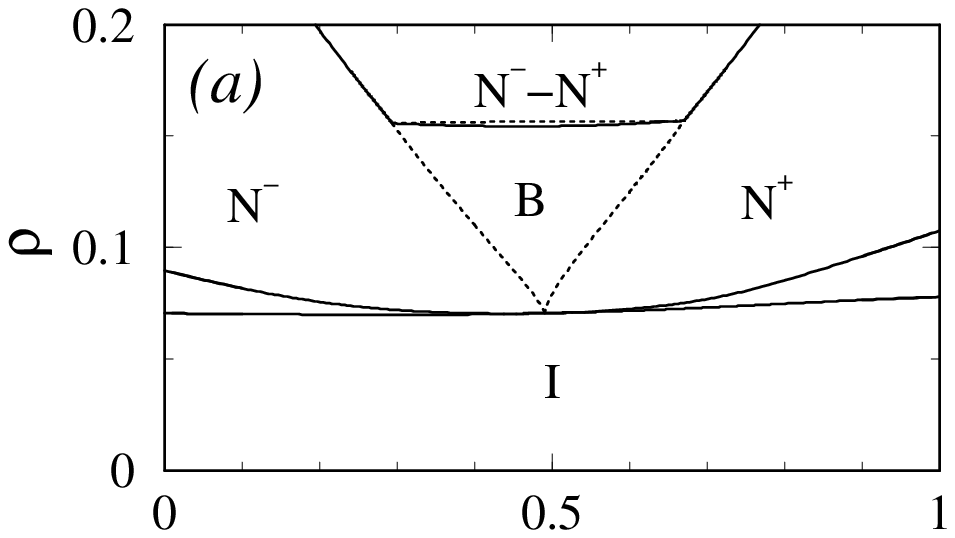}}
\hspace*{2mm}\mbox{\includegraphics*[width=3.3in, angle=0]{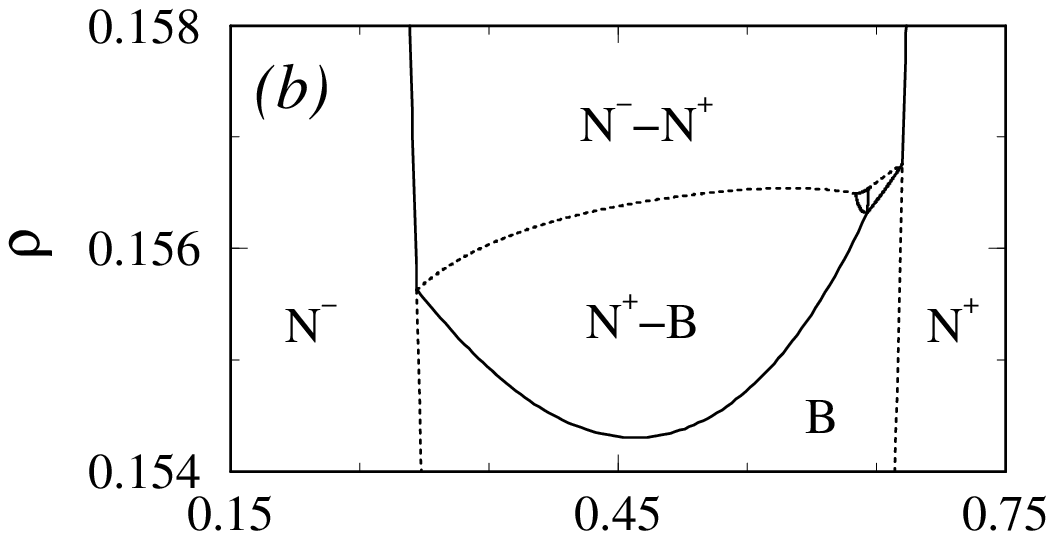}}
\mbox{\includegraphics*[width=3.4in, angle=0]{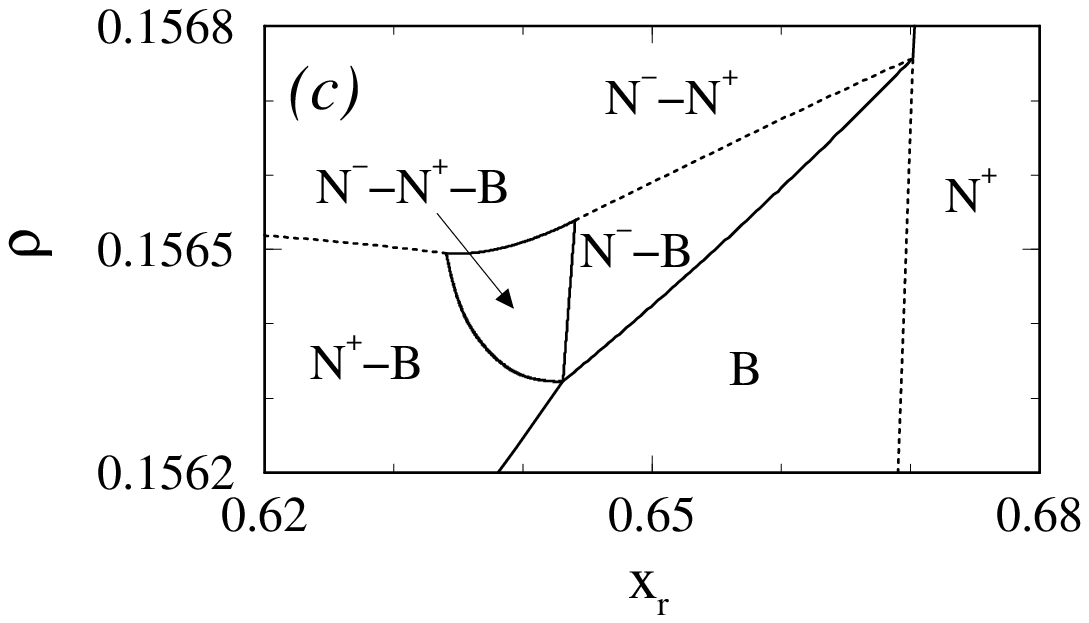}}
\caption[]{Phase diagram (global density, $\rho$, vs.\ rod molar
fraction, $x_r$) and two details of a polydisperse rod-plate mixture 
with length and breadth polydispersity $\Delta_L=0.288$ and 
$\Delta_D=0.143$, and aspect ratio
$\kappa=15$. Phases are labelled I (isotropic), N$^+$ (rod-like
nematic), N$^-$ (plate-like nematic) and B (biaxial nematic). Dotted
lines mark second-order phase transitions and full lines delimite
coexistence regions.}
\label{fig1}
\end{figure}

We have next made the mixture polydisperse. The breadth of the two
peaks in $p(\lambda)$ is controlled by $\alpha$ in (\ref{eq:parent}).
A quantitative characterization of the polydispersity can be given
if we determine the dispersion in $L$ and $D$ as obtained from 
$p(\lambda)$ for $\zeta=0$ or 1. This yields
\begin{equation}
\Delta_{L,D}\equiv\sqrt{\frac{\langle\lambda^{2\nu/3}
\rangle}{\langle\lambda^{\nu/3}\rangle^2}-1}=
\sqrt{\frac{K_{\nu/3}(\alpha)K_0(\alpha)}{K_{\nu/6}
(\alpha)^2}-1},
\end{equation}
where $\nu=2$ for $\Delta_L$  and $\nu_D=1$ for $\Delta_D$.
For $\kappa=15$ we have chosen $\alpha=1$ ($\Delta_L=0.288$, 
$\Delta_D=0.143$). The
resulting phase diagram ($\rho$ vs.\ $x_r$) is plotted in Fig.~\ref{fig1}.
The global picture (Fig.~\ref{fig1}a) resembles very much the
bidisperse case: there are first order I--N$^{\pm}$ transitions;
the two uniaxial nematics are separated by a B phase through second
order transition lines; roughly above a
threshold density the latter is replaced by a wide region of
N$^+$--N$^-$ coexistence; and there is a multicritical point where
the two I--N$^{\pm}$ first order transitions and the two N$^{\pm}$--B
second order ones meet (very close to $x_r=1/2$). There is a
difference with respect to the bidisperse phase diagram in that
the lines delimiting coexistence regions do not correspond to
coexisting states, but are the cloud lines (defined by points where 
a incipient new phase is forming) of the corresponding
coexistences. For the sake of clarity, the shadow lines (density
and composition of the incipient phase) are not represented. Notice
that $\rho$ is the global density of the system, and does not coincide
with that any of the
coexisting phases (except at the cloud lines of biphasic coexistence).
The details (Figs.~\ref{fig1}b and c) show that right above the B phase
there are regions of B--N$^{\pm}$ as well as triphasic
coexistence. Also remarkable are the two second-order transitions
separating the B--N$^{\pm}$ coexistence regions from the N$^+$--N$^-$
one, where upon increasing concentration the biaxial order
parameter vanishes continuously, hence transforming the B phase
into the second uniaxial nematic. These lines have no analogue in
the bidisperse system. Finally, a global effect of polydispersity
is to lower the phase diagram in densities and to make it more
symmetric (probably two related effects, because the lower the
density the less relevant the higher virial terms).

\begin{figure}
\hspace*{5mm}\mbox{\includegraphics*[width=3.2in, angle=0]{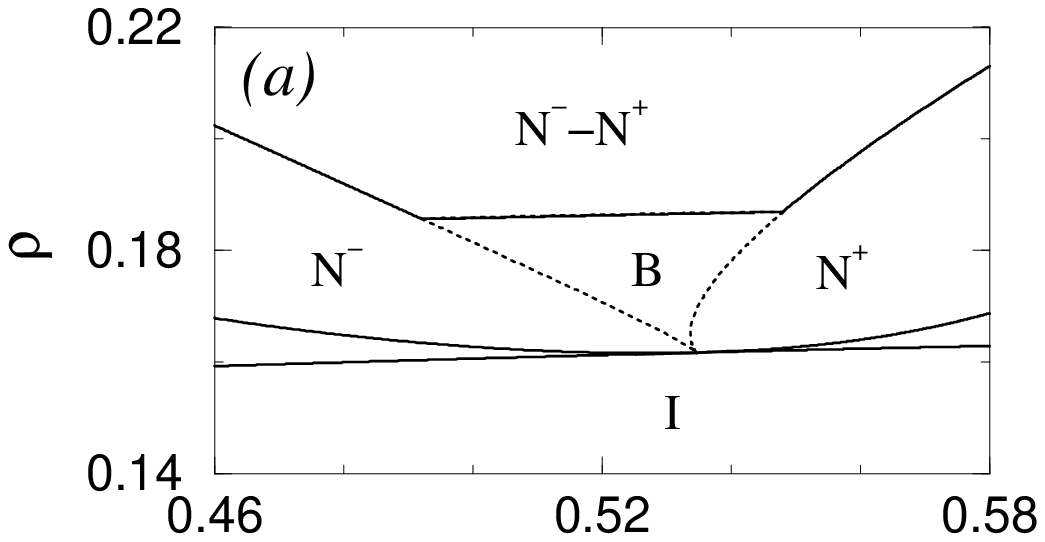}}
\mbox{\includegraphics*[width=3.4in, angle=0]{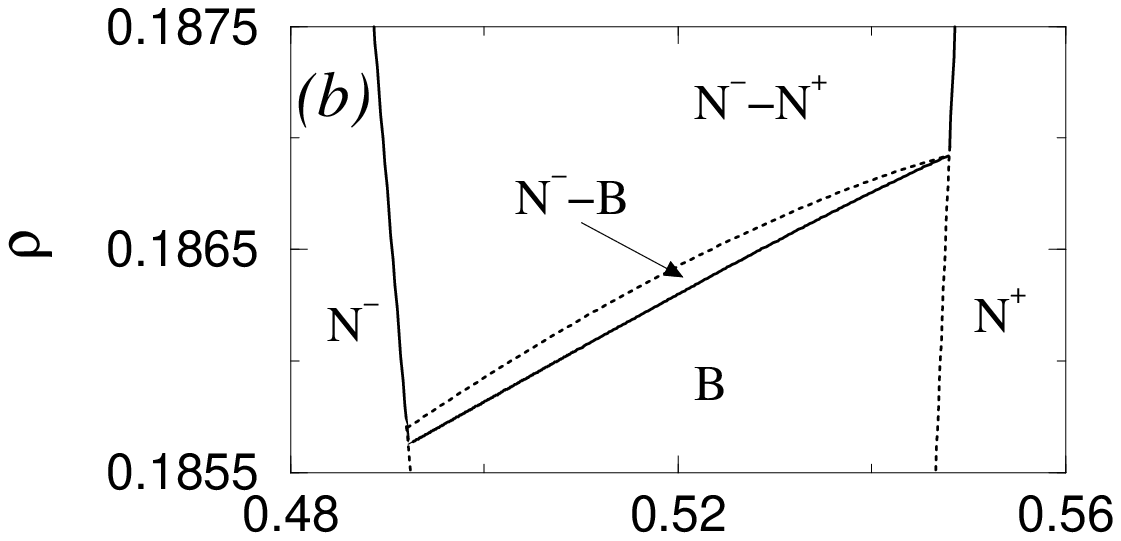}}
\caption[]{Same as Fig.~\ref{fig1} but for aspect ratio $\kappa=5$
and length and breath polydispersity $\Delta_L=0.610$ and
$\Delta_D=0.302$.}
\label{fig2}
\end{figure}

We have checked that the region of stability of the B phase is 
not appreciably affected by an increase in polydispersity 
(using $\alpha=0.1$, i.e.\ $\Delta_L=0.610$, $\Delta_D=0.302$), 
so polydispersity does not seem to inhibit
B ordering. In fact, it acts otherwise, favouring B ordering
against N$^+$--N$^-$ demixing, as the case $\kappa=5$ illustrates
(see Fig.~\ref{fig2}). Again the phase diagram of this case
looks very much like that of the bidisperse case if we take
$\alpha=1$, i.e.\ no B phase appears. However, increasing
polydispersity up to $\alpha=0.1$ we again obtain a region
where B ordering is more stable than N$^+$--N$^-$ demixing 
(Fig.~\ref{fig2}a). The B phase is limited from above by a region
of B--N$^-$ coexistence, which upon increasing density again 
becomes N$^+$--N$^-$ through a second order transition.

This enhancement of biaxial ordering is the most sriking effect
of polydispersity. To understand why it is so we must find a
mechanism by which B ordering is entropically favoured w.r.t.\
N$^+$--N$^-$ demixing. This mechanism is two-folded: on the
one hand mixing entropy increases upon increasing polydispersity,
thus penalizing demixing; but on the other hand, polydispersity
increases the gain in free volume of the B phase w.r.t.\ that
of uniaxial nematics (the average rod-plate excluded volume divided
by the average rod-rod one in a perfect B phase scales as
$\kappa^{2/3}e^{-c\Delta^2}$, $c$ being a positive constant, for 
large aspect ratios $\kappa$ and polydispersities in the
range $1\ll\Delta_{(L,D)}\ll\sqrt{\ln\kappa}$).

This effect can be exploited in the design of an experiment
to observe the B phase. The fabrication of rod-like and 
plate-like colloidal particles could follow similar procedures to
those employed by van der Kooij and Lekkerkerker 
\cite{vanderkooij:2000b}. Then, the mean size and polydispersity
of the rods and the plates can be controlled by inducing successive
fractionations (e.g.\ by adding a nonadsorbing polymer 
\cite{vanderkooij:2000a}), starting off from the appropriate parent 
distributions so as to obtain the desired final values. The key
is to produce as symmetric (in particle volume) polydisperse mixtures
as possible, since this seems a crucial ingredient for the stability
of the B phase, with or without polydispersity. How destabilizing
are the asymmetry of the mixture or the existene of a particle-volume 
distribution (motivated, for instance, by having independent 
polydispersities in length and thickness) is something that has yet 
to be quantified, but we expect these results to be robust against 
perturbations in these two directions.

A final point to be addressed is that we have not taken into account
possible nonuniform phases in the determination of the phase diagram.
According to the availble experiments \cite{vanderkooij:2000b}
these phases appear at sufficiently high concentration, while the
relevant parts of the phase diagrams we present all occur at rather low
concentrations. So it seems unlikely that these parts are affected by
the presence of inhomogeneous phases, but this has yet to be analyzed
with some care.

This work is part of the research project BFM2000-0004 (DGI) of
the Ministerio de Ciencia y Tecnolog\'{\i}a (Spain). YMR is a
postdoctoral fellow of the Consejer\'{\i}a de Educaci\'on de la 
Comunidad de Madrid (Spain).

\end{document}